\documentclass[%
superscriptaddress,
 amsmath,amssymb,
prl,
twocolumn,
]{revtex4-2}
\usepackage[export]{adjustbox}
\usepackage{graphicx}% Include figure filessn
\usepackage{color}% Include figure files
\usepackage{xcolor}% Include figure files
\usepackage{natbib}
\usepackage{wasysym}% Include figure files
\usepackage{dcolumn}% Align table columns on decimal point
\usepackage{bm}% bold math
\usepackage{hyperref}% add hypertext capabilities
\definecolor{mygreen}{rgb}{0,0.5,0}
\definecolor{mybrown}{rgb}{0.65,0.16,0.16}

\def\beq {\begin{equation}}
\def\eeq {\end{equation}}
\def\beqa {\begin{eqnarray}}
\def\eeqa {\end{eqnarray}}
\def \bnum {\begin{enumerate}}
\def \enum {\end{enumerate}}
\def\bi {\begin{itemize}}
\def\ei {\end{itemize}}

\begin{document}

\title{\noindent {\bf The saturation of exponents
and the asymptotic fourth state of turbulence}}

\author {Katepalli R. Sreenivasan} 
\affiliation {Department of Mechanical and Aerospace Engineering, New York University, New York, NY, $11201$, USA}
\affiliation {Department of Physics and the Courant Institute of Mathematical Sciences, New York University, New York, NY $11201$, USA}
\email{krs3@nyu.edu}

\author {V.Yakhot} 
\affiliation{Department of Mechanical and Aerospace Engineering, New York University, New York, NY, $11201$, USA}
\affiliation {Department of Mechanical Engineering, Boston University, Boston, MA 02215}

\begin{abstract}
A recent discovery about the inertial range of homogeneous and isotropic turbulence is the saturation of the scaling exponents $\zeta_n$ for large $n$, defined via structure functions of order $n$ as $S_{n}(r)=\overline{(\delta_r u)^{n}}=A(n)r^{\zeta_{n}}$. We focus on longitudinal structure functions for $\delta_r u$ between two positions that are $r$ apart in the same direction. In a previous paper (Phys.\ Rev.\ Fluids 6, 104604, 2021), we developed a theory for $\zeta_n$, which agrees with measurements for all $n$ for which reliable data are available, and shows saturation for large $n$. Here, we derive expressions for the probability density functions of $\delta_r u$ for four different states of turbulence, including the asymptotic fourth state corresponding to the saturation of exponents for large $n$. This saturation means that the scale separation is violated in favor of a strongly-coupled quasi-ordered flow structures, which take the form of long and thin (worm-like) structures of length $L$ and thickness $l=O(L/Re)$.
\end{abstract}
\maketitle

\paragraph*{Problem definition:} We are interested in fluid flows described by the Navier-Stokes equations 
\begin{equation}
\partial_{t}{\bf u}+{\bf u\cdot\nabla u}=-\nabla p +\nu\nabla^{2}{\bf u} + {\bf f} 
\end{equation}
with incompressibility imposed by $\nabla \cdot {\bf u}=0$. Depending on the Reynolds number, Eq.~(1) describes both laminar and turbulent flows. A goal of proper theory is to deduce the observed properties of turbulence from (1) without making {\it ad hoc} approximations. To make progress in exploring the small scale properties in the turbulent state, it is customary \cite{K41a} to introduce the velocity increments $u_r \equiv \delta_{r}u\equiv u(x+r)-u(x)$ and define structure functions as their moments. In particular, there is a natural expectation that power laws of the form
\begin{equation}
S_{n}(r)=\overline{u_{r}^{n}}=A(n)r^{\zeta_{n}}
\end{equation}
hold for intermediate separation distances within the inertial range $L \gg r \gg \eta$, where $L$ is the large scale of turbulence and the dissipation scale $\eta = (\nu^3/\varepsilon)^{1/4}$; here $\nu$ is the kinematic viscosity of the fluid and $\varepsilon$ is the rate of energy dissipation. Kolmogorov's theory \cite{K41a}, based an the assumption of locality and isotropy, led subsequently to the famous linear relation, $\zeta_n = n/3$. This elegant result invariably fails for high-Reynolds-number velocity fields in three dimensions, in which rare and extreme events dominate the tails of the probability density functions (PDFs). Increasingly intense fluctuations can be probed by considering $\langle (u_r)^n \rangle ^{1/n}$ for increasing moment order $n$, which requires the knowledge of $\zeta_{n}$ for $0\leq n \leq\infty$. Thus a major problem of the turbulence theory, similar to those of high energy and condensed matter physics, is the evaluation of the exponents $\zeta_n$ in Eq.~(2). 

\vspace{0.15cm}
\paragraph*{Specific contributions and significance:}
In \cite{krsvy}, combining Hopf equations with dynamic renormalization for the pressure gradient equations, we developed just such a theory for white-in-time Gaussian forcing at large scales, valid for all $n > 0$. We theoretically derived an expression for the scaling exponents of structure functions in isotropic turbulence, as 
\begin{equation}
\zeta_{2n}=\frac{0.366n}{0.05n+0.475}=7.32-\frac{69.54}{n+9.5},
\end{equation} 
\noindent in excellent agreement with the available data \cite{krsvy}. Here, on the basis of this theory, we distinguish four distinct states of turbulence here, and derive expressions for the PDFs of velocity increments in all these states, including what we regard as the asymptotic (or the fourth) state, which is characterized by the saturation of $\zeta_n$ with respect to $n$, as shown by (3).~This saturation immediately shows that $\langle (u_r)^n \rangle ^{1/n}$, which is characteristic of intense small scale fluctuations when $n$ is large, approaches unity quite rapidly in $n$, suggesting that small-scale fluctuations as large as the largest scale are quite prevalent --- thus making obsolete the concept of scale-separation and confirming the violation of local Galilean invariance. 

\vspace{0.15cm}
To strengthen this last point, we recall from Monin and Yaglom \cite{moni} that the Kolmogorov theory assumes turbulence in the inertial range to be universal and independent of both $L = O(1)$ and $\eta$. However, at very high Reynolds numbers (with $\nu \to 0$), the length scale $L$ remains $O(1)$ but $\eta \to 0$, becoming the dynamically important contributor to the formation of coherent, ``worm-like and pancake-like", structures, with $L$ and $\eta$ as their linear dimensions. This makes turbulence the problem of strong interactions. Indeed, if we take the only characteristic length scale in the inertial range to be the space increment $r$ and its characteristic velocity $u_{r}$, the effective local viscosity $\nu(r) \approx ru_{r}$ and the characteristic Reynolds number, by multiplying nonlinear terms of Wyld's perturbation expansion \cite{wyld}, is $R_{\lambda} = ru_{r}/\nu(r) =$ constant. A formal calculation yields this constant to be $\approx 8.8$ \cite{lesl}, implying that the perturbation theory is decidedly divergent and belongs to a class of strong interactions. Overcoming this barrier was a major contribution of \cite{krsvy}.

\vspace{0.15cm}
\paragraph*{The first state of turbulence:}
According to \cite{donz1,schu3,donz2}, in low-Reynolds number regime defined by $Re= u'L/\nu\leq 8.8$, where $u'$ is the root-mean-square velocity, the weak fluctuations generated by forcing on a large length scale $L$ reside in the scale range $r > L$ and obey Gaussian statistics. (If the forcing is different, one expects the PDF in this state of turbulence to be accordingly different.) For this condition, there is no distinction to be made between $L$ and the Taylor microscale $\lambda$, so we might as well state that $Re = R_\lambda$, where the microscale Reynolds number is based on $\lambda$ instead of $L$. This is the first stage of ``turbulence". Its character is dependent on details of forcing.

\vspace{0.15cm}
\paragraph*{The second state of turbulence:}
In statistically isotropic turbulence, if the moments of velocity increments $u_{r}$ are 
given by power laws (2), 
their  
probability density function can be found from the Mellin transform
\begin{equation}
P(u_{r},r)=\frac{1}{u_{r}}\int_{-i\infty}^{i\infty} A(n)r^{\zeta(n)}u_{r}^{-n}dn,
\end{equation}
\noindent where we have set the integral scale $L$ and the dissipation rate $\varepsilon$ equal to unity. Multiplying (4) by $u_{r}^{n}$ and evaluating the integral yields $S_{n}=A(n)r^{\zeta_{n}}$. The information available via PDFs provides a different mode for detailed comparisons with experiment and simulations; this will be discussed in a subsequent paper.

\vspace{0.15cm}
At the transition or instability point at $R_{\lambda}\approx 8.8$, this Gaussian state becomes unstable due to nonlinearity, giving rise to the smaller scale fluctuations in the interval $L> r> \eta$, making the problem much harder \cite{lesl,donz1}. The dynamics of spreading energy in the range between $L$ and $\eta$ has often been thought in terms of a cascade with constant energy flux leading to the formation of successively smaller scales, $L_{1}=L/2$, $L_{2}=L/4$, $L_{3}=L_/8$, etc., proceeding all the way to $\eta$.
It had also been assumed in early models that each step of the cascade filled the entire space. Recent work \cite{fris,SA97,KIKRS20} has shown that the above picture is flawed as a general rule.

\vspace{0.15cm}
To make use of (4), we need dynamic information on both the amplitudes $A(n)$ and the exponents $\zeta_{n}$ in (2). We obtain the expression for $A(n)$ from the large scale boundary condition for the PDF. To do this, we first have to define the scale $L$ more precisely. Based on experimental data and theoretical considerations discussed immediately below, $L$ is the scale at which the energy flux toward small scales changes sign or tends to zero. This suggests \cite{moni} that at small scales $r<L$ the structure function $S_{3}(r)<0$,
while for the larger scales $r>L$, $S_{3}\geq 0$. Typically, at this scale $L$, which depends upon the geometric details of the flow, the odd moments $S_{2n+1}(L)=0$
and the even moments saturate; i.e., $\partial _{r}S_{2n}(L)=0$. This property of turbulence has recently been examined both numerically and experimentally. The scale $L$ appears naturally in Navier-Stokes equations defined on an infinite domain driven by the white-in-time forcing function $f(k)$, $k$ being the wave number, with the variance 
$$\overline{f^{2}(k)}=\frac{{\cal P}}{2(2\pi)^{4}}\delta(k-k_{f})/k^{2}$$
\noindent where ${\cal P}$ is the forcing power and $k_f$ is the forcing wave number. The exact calculation of the relation for the third-order structure function $S_{3}(r)$ gives an oscillating expression \cite{rasm}
$$S_{3}=-{\cal P}\frac{-36r\cos r+12\sin r-12(-2+r^{2})\sin r}{r^{4}}.$$
\noindent In the limit $r\rightarrow 0$, we have the Kolomogorov-like relation 
$$S_{3}=-(4/5){\cal P}r.$$
\noindent It is interesting that no viscosity and, therefore, no dissipation scale $\eta$, appears in the above relation that resembles Kolmogorov's $4/5$-ths law, though not identical to it. At the large scale $L\approx 5.88/k_{f}$, $S_{3}(L)= 0$ \cite{fors,lesl}. In all the flows studied, the PDF $P(u_{L}, L)$ is very close to Gaussian. We would like to stress that the integral scale defined this way is not the largest scale (typically the size of the system) but, rather, corresponds to the top of the inertial range where a constant energy flux sets in toward small scales. The Gaussian boundary condition at $r=L$ is the result of fluctuation-dissipation theorem and leads to the expression $A(2n)=(2n-1)!!$ (well-tested both experimentally and numerically \cite{krsvy}).

\vspace{0.15cm}
In the second state of turbulence, we consider ``normal scaling" $\zeta_n=an$, where $a$ is a constant; we stress that $a=1/3$ is a consequence of the Kolmogorov-like relation $S_{3} \propto r$ with $S_{3}(r=0)=0$. 
Writing $(2n-1)!!=\frac{2^{n}}{\sqrt{\pi}}\int_{-\infty}^{\infty}e^{-x^{2}}x^{2n}dx$ and rotating the integration axis by $90^{o}$, 
we have
\begin{align}
P(u_r,r)=\frac{1}{\sqrt{\pi} u_r}\int_{-\infty}^{\infty}e^{-x^{2}}dx\int_{-\infty}^{\infty} e^{in( ln \frac{r^{a}\sqrt{2}}{u_r}+ ln\ x)}dn\nonumber\\
=\frac{1}{\sqrt{\pi} u_r}\int_{-\infty}^{\infty}e^{-x^{2}}\delta(ln \frac{r^{a}\sqrt{2}}{u_r}+ ln\ x)dx.
\end{align}
\noindent where $\delta$ is the standard delta function. This integral is evaluated readily to yield the result 
\begin{equation}
P(u_r)=\frac{1}{\sqrt{2 \pi}r^{a}}e^{-(\frac{u_r^{2}}{2r^{2a}})}.
\end{equation}

\noindent As we see, an infinite flow governed by (1) is characterized by a single length scale in this state. We expect these considerations to hold essentially for $n$ close to 3.

\vspace{0.15cm}
\paragraph*{The third state of turbulence:} We demonstrate the emergence of anomalous scaling by introducing small deviations from  the linear relation for $\zeta_n$, as 
\begin{equation}
\zeta_n = an-bn^{2},
\end{equation}
\noindent which, for a moment order $n$ that is not too large, can be perceived as containing the first two terms of the Taylor expansion of $\zeta_{n}$ near $n=0$. Formula (7) is not derived from the equations of motion but is a low-order expansion of $\zeta_{n}$, independent of the detailed nature of the problem. Using the Kolmogorov constraint $\zeta_{3}=1$, we get  $b=(3a-1)/9$. 
The PDF of $u_r$ is then given by
\begin{equation}
P(u_r,r)=\frac{2}{\pi u_r\sqrt{4 \ln r^{b}}}\int_{-\infty}^{\infty}e^{-x^{2}}
exp[-\frac{ (\ln \frac{u_r}{r^{a}\sqrt{2}x})^{2}}{4b\ln r}]dx.
\end{equation}
\noindent It is clear that the expansion (7) cannot be correct for all $n$. Indeed, in accordance with the H\"older inequality, $\zeta_{n}$ is a concave and non-decreasing function giving $\zeta_{n}/n\geq 1/n$ as $n\rightarrow \infty$.
So, it is amazing that for $n\leq 10$, the experimental data on strong turbulence are consistent with $a\approx 0.383$ and $b\approx 0.0166$, and that the expression (8) is accurate up to $n = O(10)$; see \cite{krsvy}. 

\vspace{0.15cm}
\paragraph*{The fourth and final state of turbulence:} This corresponds to the case of saturated $\zeta_n$. We first explore some qualitative consequences of the saturation of exponents, which do not depend on the precise saturation value; we explore that particular aspect momentarily. As indicated just below Eq.~(2), the largest fluctuations of scale $r$ have amplitudes given by $\langle u_r^n \rangle^{1/n}$ for large $n$, which, as a result of saturation of $\zeta_n$ in Eq.~(2), will have amplitudes as much as the large scale velocity itself, say $u_0$. The Reynolds number of the finest of these large fluctuations should be unity, which specifies scale $l$ via the requirement that $u_0l/\nu = 1$, giving $l = \eta Re^{-1/4}$. That is, there are very large excursions on scales that are smaller than the Kolmogorov scale $\eta$ by the factor $Re^{1/4}$, with their amplitudes of the order of $u_0$ itself. The corresponding finest time scale will also be smaller than the conventional estimate by the factor $Re^{-1/4}$. Technically, then, computationally resolving the smallest scales of motion requires grid size which is better than $\eta$ by the factor $Re^{1/4}$, and a better time resolution by the same factor of $Re^{1/4}$, than is adopted in standard direct numerical simulations (DNS) of Eq.~(1). This becomes a more conspicuous demand as the Reynolds number becomes large.

\vspace{0.15cm}
This observations blunts the unimaginable progress made by DNS in the last fifty years. The development of powerful computers has resulted in vastly larger computational domains from $N = 32^{3}$ in early '70s \cite{orsz} to $N = 16384^{3}$ today \cite{PKY}; here $N$ is the number of grid points in a periodic box within which the forced turbulence is studied. In spite of this success, it has become clear that DNS is unable to keep up because the scale ranges in both time and space widen well beyond the Kolmogorov estimates, as demonstrated just above. As a result, theoretical models based on various averaging methods and Dynamic Renormalization Group have important roles to play, especially in engineering. In particular, the one-loop renormalized perturbation expansions have led to various models for turbulent viscosity, which have been successful in simulations of the large scale flow features.

\vspace{0.15cm}
\paragraph*{The PDF of $u_r$ in the fourth state:}

\noindent To obtain the PDF, one needs the limiting value of $\zeta_n$ for $n\rightarrow\infty$; according to \cite{krsvy}, $\zeta_{n} \rightarrow 7.3$ in the limit. It should be pointed out that similar saturation properties are shared, under more ready circumstances, by the random Burgers equation \cite{yakhb,fren,becb} and the passive scalar \cite{iyer}.

\vspace{0.2cm}
The probability density function
\begin{equation}
P(u_r,r)=\frac{2}{\sqrt{\pi} u_r}\int_{-\infty}^{\infty}e^{-x^{2}}dx\int_{-i\infty}^{i\infty}x^{2n}r^{\zeta_{n}}u_r^{-n}dn
\end{equation}
\noindent can be evaluated for constant $\zeta_n$, by the use of the steepest descent approximation, yielding
\begin{equation}
P(u_r,r)\propto \frac{1}{u_r}(\frac{r^{a}}{u_r})^{\frac{1}{b}}e^{-\sqrt{|\ln u_r| |\ln r^{a}|}}.
\end{equation} 

\noindent It describes the PDF with algebraically decaying tails familiar in the literature on three-dimensional turbulence. In the limit $n\rightarrow\infty$, the PDF is qualitatively and dramatically different from (8) above, and we have
\begin{align}
P(u_r,r)=\frac{2}{\sqrt{\pi} u_r}\int_{-\infty}^{\infty}e^{-x^{2}}dx\int_{-i\infty}^{i\infty}x^{2n}r^{\zeta_{n}}u_r^{-n}dn \nonumber \\=
(\frac{r}{L})^{7.3}g(U),
\label{eq:d2}
\end{align}

\noindent where $U=2u'$ and the single point PDF is equal to 
\begin{eqnarray}
g(U)=\exp({-U^{2}/2})/\sqrt{2},
\end{eqnarray}

\noindent with $U=2u_{r}$ for $r \geq L$. Needless to say, this is applicable in the saturation state with the asymptotic exponent value of 7.3.

\vspace{0.15cm}
\paragraph*{The breakdown of local Galilean invariance in the final state:}
The nonlinearity in (1) is a consequence of Galilean invariance. Indeed, transformation to a frame moving with the velocity ${\bf u+ V }$, where ${\bf V}=$ constant, keeps the Navier-Stokes equations for fluctuations unchanged. However, it is clear from (12) that Galilean invariance is broken locally. This dynamics was experimentally studied \cite{krs98} and was shown to be possess a single-point, non-Galilean-invariant contribution to $P(u_r,r; U)$. This breakdown of Galielean invariance had been previously assumed \cite{poly} in the context of the Burgers equation (which, as already pointed out, also possesses saturated scaling exponents). We would like to stress that saturation emerged in \cite{krsvy} as a solution to the Hopf equation. One can hardly expect universality in this regime. 

\vspace{0.15cm}
\paragraph*{Geometric structure:} The analysis so far demonstrates the relevance of both large and small scales when fluctuations are intense. This points to the existence of powerful structures one of whose dimensions is very small, of the order $l = \eta Re^{-1/4}$, and the other of the order $L$. Between vortex sheets and tubes, it would appear that the inherent instability of the sheets and their tendency to roll up suggests that the final structures are likely to be in the form of tubes. If tubes are the most likely objects, it is clear that they are like the ``worms" described in many simulations, see especially \cite{kaneda}, with length of the order $L$. These vortical movements, which are very thin and long at the same time, typically covering substantial fractions of the flow field, are a feature of all high-Reynolds-number flows such as homogeneous turbulence, thermal convection, meteorological flows, etc., and are more readily observed with passing time. 

\vspace{0.15cm}
Given this picture of high-Reynolds-number turbulence, it is clear that no local filtering procedure can be applied successfully. Neither Kolmogorov-like arguments nor other qualitative or approximate approaches can account for structures that are very small and very large simultaneously. It might have been reasonable to do so if they were extremely rare, but the saturation of exponents makes them not so rare. In particular, this property of the asymptotic state does not augur well for large-eddy-simulation methods.

\vspace{0.15cm}
\paragraph*{Summary and conclusions:} 
 
\noindent In our previous paper \cite{krsvy}, we considered the dynamics of spatially infinite fluid driven by white-in-time Gaussian random force acting on a length scale $L$. Our main interest was to probe the velocity field in terms of structure functions defined in (2). Of particular interest is $(\langle u_r^{n} \rangle)^{1/n}$ for large $n$, the magnitude of local mean velocity corresponding to the far tails of the PDFs, dominated by extreme events. 

\vspace{0.15cm}
We noted that many years of experimental and numerical work \cite{fris,SA97,KIKRS20} has revealed power laws
with anomalous exponents departing from $\zeta_{n}=n/3$ (except for $n=3)$. The theory of turbulence, briefly mentioned in this paper, giving the result (3), was produced only recently \cite{krsvy}. Here, based on that exact expression for the exponents, we have analyzed the PDFs of velocity increments to investigate the structure of turbulence. To estimate the effect of intermittency in the limit of large $n$, we write 
\begin{equation}
\Gamma(n)=\frac{\zeta_{n}^{41}}{\zeta_{n}}=0.0455n+0.864.
\end{equation}  
\noindent When $n \le 4$, this ratio is within a few percent of unity, showing that intermittency is practically not important in this range. However, as $n\rightarrow \infty$ (say, for $n\geq 20$), the intermittent behavior dominates the velocity field --- meaning that all models based on arguments of \cite{K41a}, which do not recognize intermittency, are quite problematic.

\vspace{0.15cm}
We can now present the basic ingredients of the turbulence theory, valid in the entire range of Reynolds numbers, as follows: \\
   
{\it 1. Linear regime:} The velocity $\bf u$ is proportional to the forcing $\bf f$. This corresponds to the lowest Reynolds number, $R_{\lambda}\rightarrow 0$ \cite{krsvy}. 

{\it 2. The low Reynolds number range, $R_{\lambda}< 8.8$:} The quasi-laminar Gaussian random flow consists of random patches of the typical scale $L$ \cite{donz1,schu3,donz2}. In this range, $L$ is the only relevant scale, and, in particular, it is of the same order as $\lambda$. 

{\it 3. Range of anomalous scaling:} At $R_{\lambda}\approx 8.8$ local transition to anomalous scaling ensues, seen by the broadening of the PDF tails of velocity increments. %In \cite{donz1,schu3,donz2}, this was also accompanied by the appearance of sparse spots with local $R_{\lambda}>8.8$.

{\it 4. Well-defined structures:} With further increase of $R_{\lambda}$, the tails of the PDFs broaden and the top becomes sharper \cite{krs}. We have argued that this is accompanied by the appearance of elongated, thin and relatively rare structures responsible for strong intermittency. 

{\it 5. The asymptotic state:} In the final or limiting state of turbulence, one might regard the entire flow as a Gaussian ${\it gas}$ of long spaghetti-like structures with length of order $L$ and thickness $l = O(L/Re)$. 

\vspace{0.15cm}
We conclude on an optimistic note: The theory developed in \cite{krsvy} and in the present paper can be generalized to the cases of compressible hydrodynamic turbulence and gravitational  collapse of interstellar dust which are of interest in processes of star formation and related applications. They will be the topic of future studies.  

\section{Acknowledgments}
We have interacted with many colleagues too numerous to list here, but acknowledge the early influence of the late S. Orszag, and discussions with H. Chen, D.A. Donzis, G.L. Eyink, K.P. Iyer, A.A. Migdal, A.M. Polyakov, J. Schumacher, L. Smith, and I. Starosel'sky.

\end{document}